\documentclass[a4paper,12pt]{article}
\usepackage{amsfonts,amsmath,amssymb,amscd,amsthm,latexsym,url}

\newcommand{\bi}{\bibitem}
\newcommand{\nb}{\newblock}
\newcommand{\be}[1]{\begin{equation}\label{#1}}
\newcommand{\ee}{\end{equation}}

\newtheorem{thm}{\quad Theorem}

\begin{document}
\title{Polynomial time algorithm for 3-SAT. Examples of use.}%
\author{Sergey Gubin}%
\thanks{Author's email: sgubin@genesyslab.com}
\begin{abstract}
The algorithm checks the propositional formulas for patterns of unsatisfiability.
\end{abstract}
\maketitle

\section{Introduction}
An $O(m^{3})$-time algorithm for 3-SAT is described in \cite{gubin}. This article illustrates that work with a few examples.
\newline\indent
Let's summarize the algorithm. Let $c_{1}, c_{2}, \ldots, c_{m}$ be the clauses of a given 3-SAT instance. The algorithm detects whether there are compatible truth assignments among those truth assignments, which satisfy separate clauses.
\newline
\emph{Start.} Based on the clauses' truth-tables, let's build the clauses' compatibility matrices, grouped in the following box-matrix:
\[
\begin{array}{cccc}
c_{1} : c_{2} & c_{1} : c_{3} &  \ldots &c_{1} : c_{m} \\
						  & c_{2} : c_{3} &  \ldots &c_{2} : c_{m} \\
						  &               &  \ddots & \vdots \\	
						  &               &  \ldots &c_{m-1} : c_{m} \\						  					  
\end{array}
\] 
\emph{Step 1.} Based on the first row of the box-matrix of compatibility matrices, let's deplete the compatibility matrices in the rows starting from the second. The resulting matrix is the following:
\[
\begin{array}{cccc}
c_{1} : c_{2} & c_{1} : c_{3} &  \ldots &c_{1} : c_{m} \\
						  & c_{1} \wedge c_{2} : c_{1} \wedge c_{3} &  \ldots & c_{1} \wedge c_{2} : c_{1} \wedge c_{m} \\
						  &               &  \ddots & \vdots \\	
						  &               &  \ldots &c_{1} \wedge c_{m-1} : c_{1} \wedge c_{m} \\						  					  
\end{array}
\] 
\emph{Next steps.} Based on the second row of the resulting matrix, let's deplete the compatibility matrices in the rows starting from the third. And so on. After $m-2$ steps the matrix in the last row of the box-matrix of the compatibility matrices gets to be
\[
c_{1} \wedge c_{2} \wedge \ldots \wedge c_{m-1} : c_{1} \wedge c_{2} \wedge \ldots \wedge c_{m}
\]
Thus, the given clauses are compatible (the given 3-SAT instance is satisfiable) iff the algorithm does not create matrices filled with $false$ completely.
\newline\indent
In the tables below, ``1'' is used for $true$ and empty string for $false$. Some extra white space in the article is due to allocation of the tables.

\section{Example 1}
The 3-SAT instance from \cite{cook}:
\[
f = (p \vee q \vee r) \wedge (\bar{p} \vee q \vee \bar{r}) \wedge (p \vee \bar{q} \vee s) \wedge (\bar{p} \vee \bar{r} \vee \bar{s}).
\]
Let's use the algorithm directly as it was described in Theorem 1 in \cite{gubin}. Clauses' truth-tables:
\[
c_{1} = 
\begin{array}{||c||c|c|c||c||}
\hline\hline
\#&p&q&r&p \vee q \vee r \\
\hline\hline
1& & & &  \\
\hline
2& & &1&1 \\
\hline
3& &1& &1 \\
\hline
4& &1&1&1 \\
\hline
5&1& & &1 \\
\hline
6&1& &1&1 \\
\hline
7&1&1& &1 \\
\hline
8&1&1&1&1 \\
\hline\hline
\end{array}
~ ~
c_{2} = 
\begin{array}{||c||c|c|c||c||}
\hline\hline
\#&p&q&r&\bar{p} \vee q \vee \bar{r} \\
\hline\hline
1& & & &1 \\
\hline
2& & &1&1 \\
\hline
3& &1& &1 \\
\hline
4& &1&1&1 \\
\hline
5&1& & &1 \\
\hline
6&1& &1&  \\
\hline
7&1&1& &1 \\
\hline
8&1&1&1&1 \\
\hline\hline
\end{array}
\]
\[
c_{3} = 
\begin{array}{||c||c|c|c||c||}
\hline\hline
\#&p&q&s&p \vee \bar{q} \vee s \\
\hline\hline
1& & & &1 \\
\hline
2& & &1&1 \\
\hline
3& &1& &  \\
\hline
4& &1&1&1 \\
\hline
5&1& & &1 \\
\hline
6&1& &1&1 \\
\hline
7&1&1& &1 \\
\hline
8&1&1&1&1 \\
\hline\hline
\end{array}
~ ~
c_{4} = 
\begin{array}{||c||c|c|c||c||}
\hline\hline
\#&p&r&s&\bar{p} \vee \bar{r} \vee \bar{s} \\
\hline\hline
1& & & &1 \\
\hline
2& & &1&1 \\
\hline
3& &1& &1 \\
\hline
4& &1&1&1 \\
\hline
5&1& & &1 \\
\hline
6&1& &1&1 \\
\hline
7&1&1& &1 \\
\hline
8&1&1&1&  \\
\hline\hline
\end{array}
\]
\emph{Start.} Based on the truth-tables, we calculate clauses' compatibility matrices. Compatibility matrix $c_{i} : c_{j}$ shows which meanings of clauses $c_{i}$ and $c_{j}$ are compatible, $i < j; ~ i = 1,2,3;~ j= 2,3,4$:
\[
\begin{array}{|c|c|c|}
\hline
c_{1} : c_{2} & c_{1} : c_{3} & c_{1} : c_{4} \\
\hline
						  & c_{2} : c_{3} & c_{2} : c_{4} \\
\hline						  
						  &               & c_{3} : c_{4} \\						  					  
\hline						  
\end{array}
\] 
The matrices are shown below.
\newline
\emph{Step 1.} Let's deplete compatibility matrices $c_{i} : c_{j}, ~ i \geq 2$. These matrices are allocated in the second and the third rows on the pictures below. Let's get rid of those elements $(\alpha\beta)$ of these matrices, which are not compatible with any truth assignment for clause $c_{1}$:
\[
\begin{array}{|c|c|c|}
\hline
c_{1} : c_{2} & c_{1} : c_{3} & c_{1} : c_{4} \\
\hline
						  & c_{1} \wedge c_{2} : c_{1} \wedge c_{3} & c_{1} \wedge c_{2} : c_{1} \wedge c_{4} \\
\hline
						  &               & c_{1} \wedge c_{3} : c_{1} \wedge c_{4} \\	
\hline						  					  					  
\end{array}
\] 
For example, element (1,1) of compatibility matrix $c_{3}:c_{4}$ is associated with the first meaning of clause $c_{3}$ and the first meaning of clause $c_{4}$. The numbers are taken in accordance with truth-tables for the clauses. But there is not any meaning of clause $c_{1}$, which would be compatible simultaneously with the first meaning of clause $c_{3}$ and the first meaning of clause $c_{4}$. Thus, element (1,1) of compatibility matrix $c_{3} : c_{4}$ must be eliminated.
\newline\indent
The resulting matrices are shown below. Because there is not any $false$-matrix, let's continue.
\newline
\emph{Step 2.} Let's deplete compatibility matrix $c_{3} : c_{4}$. The matrix is allocated in the third row. Let's get rid of those elements $(\alpha\beta)$, for which there is not any truth assignment for clause $c_{2}$ left, which would be simultaneously compatible with $\alpha$-th truth assignment for $c_{3}$ and with $\beta$-th truth assignment for clause $c_{4.}$. Let's mention that, after step 1, matrix $c_{3} : c_{4}$ contains only those couples, which are compatible with clause $c_{1}$:
\[
\begin{array}{|c|c|c|}
\hline
c_{1} : c_{2} & c_{1} : c_{3} & c_{1} : c_{4} \\
\hline
						  & c_{1} \wedge c_{2} : c_{1} \wedge c_{3} & c_{1} \wedge c_{2} : c_{1} \wedge c_{4} \\
\hline
						  &               & c_{1} \wedge c_{2} \wedge c_{3} : c_{1} \wedge c_{2} \wedge c_{4} \\						  	
\hline						  				  
\end{array}
\] 
The resulting matrix is shown below. Because it is not a $false$-matrix, the example is a satisfiable 3-SAT instance.
\newpage
\emph{Start.} Clauses' compatibility matrices:
\[
\begin{array}{cccc}
&c_{2}&c_{3}&c_{4} \\
c_{1}&
\begin{array}{||c||c|c|c|c|c|c|c|c||}
\hline\hline
&1&2&3&4&5&6&7&8 \\
\hline\hline
1& & & & & & & &  \\
\hline
2& &1& & & & & &  \\
\hline
3& & &1& & & & &  \\
\hline
4& & & &1& & & &  \\
\hline
5& & & & &1& & &  \\
\hline
6& & & & & & & &  \\
\hline
7& & & & & & &1&  \\
\hline
8& & & & & & & &1 \\
\hline\hline
\end{array}
&
\begin{array}{||c|c|c|c|c|c|c|c||}
\hline\hline
1&2&3&4&5&6&7&8 \\
\hline\hline
 & & & & & & &  \\
\hline
1&1& & & & & &  \\
\hline
 & & &1& & & &  \\
\hline
 & & &1& & & &  \\
\hline
 & & & &1&1& &  \\
\hline
 & & & &1&1& &  \\
\hline
 & & & & & &1&1 \\
\hline
 & & & & & &1&1 \\
\hline\hline
\end{array}
&
\begin{array}{||c|c|c|c|c|c|c|c||}
\hline\hline
1&2&3&4&5&6&7&8 \\
\hline\hline
 & & & & & & &  \\
\hline
 & &1&1& & & &  \\
\hline
1&1& & & & & &  \\
\hline
 & &1&1& & & &  \\
\hline
 & & & &1&1& &  \\
\hline
 & & & & & &1&  \\
\hline
 & & & &1&1& &  \\
\hline
 & & & & & &1&  \\
\hline\hline
\end{array}
\end{array}
\]
\[
\begin{array}{crcc}
& &c_{3}&c_{4} \\
\begin{array}{cccccccc}
~~~~~~&~ &~ &~ &~ &~ &~ &~  \\
 & & & & & & &  \\
 & & & & & & &  \\
 & & & & & & &  \\
 & & & & & & &  \\
 & & & & & & &  \\
 & & & & & & &  \\
 & & & & & & &  \\
 & & & & & & &  \\
\end{array}
&c_{2}&
\begin{array}{||c||c|c|c|c|c|c|c|c||}
\hline\hline
&1&2&3&4&5&6&7&8 \\
\hline\hline
1&1&1& & & & & &  \\
\hline
2&1&1& & & & & &  \\
\hline
3& & & &1& & & &  \\
\hline
4& & & &1& & & &  \\
\hline
5& & & & &1&1& &  \\
\hline
6& & & & & & & &  \\
\hline
7& & & & & & &1&1 \\
\hline
8& & & & & & &1&1 \\
\hline\hline
\end{array}
&
\begin{array}{||c|c|c|c|c|c|c|c||}
\hline\hline
1&2&3&4&5&6&7&8 \\
\hline\hline
1&1& & & & & &  \\
\hline
 & &1&1& & & &  \\
\hline
1&1& & & & & &  \\
\hline
 & &1&1& & & &  \\
\hline
 & & & &1&1& &  \\
\hline
 & & & & & & &  \\
\hline
 & & & &1&1& &  \\
\hline
 & & & & & &1&  \\
\hline\hline
\end{array}
\end{array}
\]
\[
\begin{array}{crcc}
& & &c_{4} \\
\begin{array}{cccccccc}
~~~~~~&~ &~ &~ &~ &~ &~ &~  \\
 & & & & & & &  \\
 & & & & & & &  \\
 & & & & & & &  \\
 & & & & & & &  \\
 & & & & & & &  \\
 & & & & & & &  \\
 & & & & & & &  \\
 & & & & & & &  \\
\end{array}
&
\begin{array}{cccccccc}
~~~~~~&~ &~ &~ &~ &~ &~ &~  \\
 & & & & & & &  \\
 & & & & & & &  \\
 & & & & & & &  \\
 & & & & & & &  \\
 & & & & & & &  \\
 & & & & & & &  \\
 & & & & & & &  \\
 & & & & & & &  \\
\end{array}
&c_{3}&
\begin{array}{||c||c|c|c|c|c|c|c|c||}
\hline\hline
&1&2&3&4&5&6&7&8 \\
\hline\hline
1&1& &1& & & & &  \\
\hline
2& &1& &1& & & &  \\
\hline
3& & & & & & & &  \\
\hline
4& &1& &1& & & &  \\
\hline
5& & & & &1& &1&  \\
\hline
6& & & & & &1& &  \\
\hline
7& & & & &1& &1&  \\
\hline
8& & & & & &1& &  \\
\hline\hline
\end{array}
\end{array}
\]

\newpage
\emph{Step 1.} Compatibility matrices in the second and the third rows are depleted by clause $c_{1}$:
\[
\begin{array}{cccc}
&c_{2}&c_{3}&c_{4} \\
c_{1}&
\begin{array}{||c||c|c|c|c|c|c|c|c||}
\hline\hline
&1&2&3&4&5&6&7&8 \\
\hline\hline
1& & & & & & & &  \\
\hline
2& &1& & & & & &  \\
\hline
3& & &1& & & & &  \\
\hline
4& & & &1& & & &  \\
\hline
5& & & & &1& & &  \\
\hline
6& & & & & & & &  \\
\hline
7& & & & & & &1&  \\
\hline
8& & & & & & & &1 \\
\hline\hline
\end{array}
&
\begin{array}{||c|c|c|c|c|c|c|c||}
\hline\hline
1&2&3&4&5&6&7&8 \\
\hline\hline
 & & & & & & &  \\
\hline
1&1& & & & & &  \\
\hline
 & & &1& & & &  \\
\hline
 & & &1& & & &  \\
\hline
 & & & &1&1& &  \\
\hline
 & & & &1&1& &  \\
\hline
 & & & & & &1&1 \\
\hline
 & & & & & &1&1 \\
\hline\hline
\end{array}
&
\begin{array}{||c|c|c|c|c|c|c|c||}
\hline\hline
1&2&3&4&5&6&7&8 \\
\hline\hline
 & & & & & & &  \\
\hline
 & &1&1& & & &  \\
\hline
1&1& & & & & &  \\
\hline
 & &1&1& & & &  \\
\hline
 & & & &1&1& &  \\
\hline
 & & & & & &1&  \\
\hline
 & & & &1&1& &  \\
\hline
 & & & & & &1&  \\
\hline\hline
\end{array}
\end{array}
\]
\[
\begin{array}{crcc}
& &c_{3}&c_{4} \\
\begin{array}{cccccccc}
~~~~~~&~ &~ &~ &~ &~ &~ &~  \\
 & & & & & & &  \\
 & & & & & & &  \\
 & & & & & & &  \\
 & & & & & & &  \\
 & & & & & & &  \\
 & & & & & & &  \\
 & & & & & & &  \\
 & & & & & & &  \\
\end{array}
&c_{2}&
\begin{array}{||c||c|c|c|c|c|c|c|c||}
\hline\hline
&1&2&3&4&5&6&7&8 \\
\hline\hline
1& & & & & & & &  \\
\hline
2&1&1& & & & & &  \\
\hline
3& & & &1& & & &  \\
\hline
4& & & &1& & & &  \\
\hline
5& & & & &1&1& &  \\
\hline
6& & & & & & & &  \\
\hline
7& & & & & & &1&1 \\
\hline
8& & & & & & &1&1 \\
\hline\hline
\end{array}
&
\begin{array}{||c|c|c|c|c|c|c|c||}
\hline\hline
1&2&3&4&5&6&7&8 \\
\hline\hline
 & & & & & & &  \\
\hline
 & &1&1& & & &  \\
\hline
1&1& & & & & &  \\
\hline
 & &1&1& & & &  \\
\hline
 & & & &1&1& &  \\
\hline
 & & & & & & &  \\
\hline
 & & & &1&1& &  \\
\hline
 & & & & & &1&  \\
\hline\hline
\end{array}
\end{array}
\]
\[
\begin{array}{crcc}
& & &c_{4} \\
\begin{array}{cccccccc}
~~~~~~&~ &~ &~ &~ &~ &~ &~  \\
 & & & & & & &  \\
 & & & & & & &  \\
 & & & & & & &  \\
 & & & & & & &  \\
 & & & & & & &  \\
 & & & & & & &  \\
 & & & & & & &  \\
 & & & & & & &  \\
\end{array}
&
\begin{array}{cccccccc}
~~~~~~&~ &~ &~ &~ &~ &~ &~  \\
 & & & & & & &  \\
 & & & & & & &  \\
 & & & & & & &  \\
 & & & & & & &  \\
 & & & & & & &  \\
 & & & & & & &  \\
 & & & & & & &  \\
 & & & & & & &  \\
\end{array}
&c_{3}&
\begin{array}{||c||c|c|c|c|c|c|c|c||}
\hline\hline
&1&2&3&4&5&6&7&8 \\
\hline\hline
1& & &1& & & & &  \\
\hline
2& & & &1& & & &  \\
\hline
3& & & & & & & &  \\
\hline
4& &1& &1& & & &  \\
\hline
5& & & & &1& &1&  \\
\hline
6& & & & & &1& &  \\
\hline
7& & & & &1& &1&  \\
\hline
8& & & & & &1& &  \\
\hline\hline
\end{array}
\end{array}
\]
\newpage
\emph{Step 2.} Compatibility matrix $c_{3} : c_{4}$ in the third row is depleted by clauses $c_{1}$ and $c_{2}$:
\[
\begin{array}{cccc}
&c_{2}&c_{3}&c_{4} \\
c_{1}&
\begin{array}{||c||c|c|c|c|c|c|c|c||}
\hline\hline
&1&2&3&4&5&6&7&8 \\
\hline\hline
1& & & & & & & &  \\
\hline
2& &1& & & & & &  \\
\hline
3& & &1& & & & &  \\
\hline
4& & & &1& & & &  \\
\hline
5& & & & &1& & &  \\
\hline
6& & & & & & & &  \\
\hline
7& & & & & & &1&  \\
\hline
8& & & & & & & &1 \\
\hline\hline
\end{array}
&
\begin{array}{||c|c|c|c|c|c|c|c||}
\hline\hline
1&2&3&4&5&6&7&8 \\
\hline\hline
 & & & & & & &  \\
\hline
1&1& & & & & &  \\
\hline
 & & &1& & & &  \\
\hline
 & & &1& & & &  \\
\hline
 & & & &1&1& &  \\
\hline
 & & & &1&1& &  \\
\hline
 & & & & & &1&1 \\
\hline
 & & & & & &1&1 \\
\hline\hline
\end{array}
&
\begin{array}{||c|c|c|c|c|c|c|c||}
\hline\hline
1&2&3&4&5&6&7&8 \\
\hline\hline
 & & & & & & &  \\
\hline
 & &1&1& & & &  \\
\hline
1&1& & & & & &  \\
\hline
 & &1&1& & & &  \\
\hline
 & & & &1&1& &  \\
\hline
 & & & & & &1&  \\
\hline
 & & & &1&1& &  \\
\hline
 & & & & & &1&  \\
\hline\hline
\end{array}
\end{array}
\]
\[
\begin{array}{crcc}
& &c_{3}&c_{4} \\
\begin{array}{cccccccc}
~~~~~~&~ &~ &~ &~ &~ &~ &~  \\
 & & & & & & &  \\
 & & & & & & &  \\
 & & & & & & &  \\
 & & & & & & &  \\
 & & & & & & &  \\
 & & & & & & &  \\
 & & & & & & &  \\
 & & & & & & &  \\
\end{array}
&c_{2}&
\begin{array}{||c||c|c|c|c|c|c|c|c||}
\hline\hline
&1&2&3&4&5&6&7&8 \\
\hline\hline
1& & & & & & & &  \\
\hline
2&1&1& & & & & &  \\
\hline
3& & & &1& & & &  \\
\hline
4& & & &1& & & &  \\
\hline
5& & & & &1&1& &  \\
\hline
6& & & & & & & &  \\
\hline
7& & & & & & &1&1 \\
\hline
8& & & & & & &1&1 \\
\hline\hline
\end{array}
&
\begin{array}{||c|c|c|c|c|c|c|c||}
\hline\hline
1&2&3&4&5&6&7&8 \\
\hline\hline
 & & & & & & &  \\
\hline
 & &1&1& & & &  \\
\hline
1&1& & & & & &  \\
\hline
 & &1&1& & & &  \\
\hline
 & & & &1&1& &  \\
\hline
 & & & & & & &  \\
\hline
 & & & &1&1& &  \\
\hline
 & & & & & &1&  \\
\hline\hline
\end{array}
\end{array}
\]
\[
\begin{array}{crcc}
& & &c_{4} \\
\begin{array}{cccccccc}
~~~~~~&~ &~ &~ &~ &~ &~ &~  \\
 & & & & & & &  \\
 & & & & & & &  \\
 & & & & & & &  \\
 & & & & & & &  \\
 & & & & & & &  \\
 & & & & & & &  \\
 & & & & & & &  \\
 & & & & & & &  \\
\end{array}
&
\begin{array}{cccccccc}
~~~~~~&~ &~ &~ &~ &~ &~ &~  \\
 & & & & & & &  \\
 & & & & & & &  \\
 & & & & & & &  \\
 & & & & & & &  \\
 & & & & & & &  \\
 & & & & & & &  \\
 & & & & & & &  \\
 & & & & & & &  \\
\end{array}
&c_{3}&
\begin{array}{||c||c|c|c|c|c|c|c|c||}
\hline\hline
&1&2&3&4&5&6&7&8 \\
\hline\hline
1& & &1& & & & &  \\
\hline
2& & & &1& & & &  \\
\hline
3& & & & & & & &  \\
\hline
4& &1& &1& & & &  \\
\hline
5& & & & &1& & &  \\
\hline
6& & & & & &1& &  \\
\hline
7& & & & &1& &1&  \\
\hline
8& & & & & &1& &  \\
\hline\hline
\end{array}
\end{array}
\]
Because there is not any $false$-matrix, the 3-SAT instance is satisfiable.
\newpage
\emph{All solutions.} In the same way, let's move backward and get rid of all elements in the box-matrix of the compatibility matrices except those satisfying the given formula:
\[
\begin{array}{cccc}
&c_{2}&c_{3}&c_{4} \\
c_{1}&
\begin{array}{||c||c|c|c|c|c|c|c|c||}
\hline\hline
&1&2&3&4&5&6&7&8 \\
\hline\hline
1& & & & & & & &  \\
\hline
2& &1& & & & & &  \\
\hline
3& & &1& & & & &  \\
\hline
4& & & &1& & & &  \\
\hline
5& & & & &1& & &  \\
\hline
6& & & & & & & &  \\
\hline
7& & & & & & &1&  \\
\hline
8& & & & & & & &1 \\
\hline\hline
\end{array}
&
\begin{array}{||c|c|c|c|c|c|c|c||}
\hline\hline
1&2&3&4&5&6&7&8 \\
\hline\hline
 & & & & & & &  \\
\hline
1&1& & & & & &  \\
\hline
 & & &1& & & &  \\
\hline
 & & &1& & & &  \\
\hline
 & & & &1&1& &  \\
\hline
 & & & & & & &  \\
\hline
 & & & & & &1&1 \\
\hline
 & & & & & &1&  \\
\hline\hline
\end{array}
&
\begin{array}{||c|c|c|c|c|c|c|c||}
\hline\hline
1&2&3&4&5&6&7&8 \\
\hline\hline
 & & & & & & &  \\
\hline
 & &1&1& & & &  \\
\hline
 &1& & & & & &  \\
\hline
 & & &1& & & &  \\
\hline
 & & & &1&1& &  \\
\hline
 & & & & & & &  \\
\hline
 & & & &1&1& &  \\
\hline
 & & & & & &1&  \\
\hline\hline
\end{array}
\end{array}
\]
\[
\begin{array}{crcc}
& &c_{3}&c_{4} \\
\begin{array}{cccccccc}
~~~~~~&~ &~ &~ &~ &~ &~ &~  \\
 & & & & & & &  \\
 & & & & & & &  \\
 & & & & & & &  \\
 & & & & & & &  \\
 & & & & & & &  \\
 & & & & & & &  \\
 & & & & & & &  \\
 & & & & & & &  \\
\end{array}
&c_{2}&
\begin{array}{||c||c|c|c|c|c|c|c|c||}
\hline\hline
&1&2&3&4&5&6&7&8 \\
\hline\hline
1& & & & & & & &  \\
\hline
2&1&1& & & & & &  \\
\hline
3& & & &1& & & &  \\
\hline
4& & & &1& & & &  \\
\hline
5& & & & &1&1& &  \\
\hline
6& & & & & & & &  \\
\hline
7& & & & & & &1&1 \\
\hline
8& & & & & & &1&  \\
\hline\hline
\end{array}
&
\begin{array}{||c|c|c|c|c|c|c|c||}
\hline\hline
1&2&3&4&5&6&7&8 \\
\hline\hline
 & & & & & & &  \\
\hline
 & &1&1& & & &  \\
\hline
 &1& & & & & &  \\
\hline
 & & &1& & & &  \\
\hline
 & & & &1&1& &  \\
\hline
 & & & & & & &  \\
\hline
 & & & &1&1& &  \\
\hline
 & & & & & &1&  \\
\hline\hline
\end{array}
\end{array}
\]
\[
\begin{array}{crcc}
& & &c_{4} \\
\begin{array}{cccccccc}
~~~~~~&~ &~ &~ &~ &~ &~ &~  \\
 & & & & & & &  \\
 & & & & & & &  \\
 & & & & & & &  \\
 & & & & & & &  \\
 & & & & & & &  \\
 & & & & & & &  \\
 & & & & & & &  \\
 & & & & & & &  \\
\end{array}
&
\begin{array}{cccccccc}
~~~~~~&~ &~ &~ &~ &~ &~ &~  \\
 & & & & & & &  \\
 & & & & & & &  \\
 & & & & & & &  \\
 & & & & & & &  \\
 & & & & & & &  \\
 & & & & & & &  \\
 & & & & & & &  \\
 & & & & & & &  \\
\end{array}
&c_{3}&
\begin{array}{||c||c|c|c|c|c|c|c|c||}
\hline\hline
&1&2&3&4&5&6&7&8 \\
\hline\hline
1& & &1& & & & &  \\
\hline
2& & & &1& & & &  \\
\hline
3& & & & & & & &  \\
\hline
4& &1& &1& & & &  \\
\hline
5& & & & &1& & &  \\
\hline
6& & & & & &1& &  \\
\hline
7& & & & &1& &1&  \\
\hline
8& & & & & &1& &  \\
\hline\hline
\end{array}
\end{array}
\]

\newpage
\emph{Brute force method.} Let's compare the solution with the solution obtained with the brute force method. The brute force solution is shown in the table below.
\[
\begin{array}{||c||c|c|c|c||c|c|c|c||c||c|c|c||c|c||c||}
\hline\hline
\#&p&q&r&s&c_{1}&c_{2}&c_{3}&c_{4}&f&c_{1}:c_{2}&c_{1}:c_{3}&c_{1}:c_{4}&c_{2}:c_{3}&c_{2}:c_{4}&c_{3}:c_{4} \\
\hline\hline
1& & & & & &1&1&1& & & & & & & \\
\hline
2& & & &1& &1&1&1& & & & & & & \\
\hline
3& & &1& &1&1&1&1&1&2:2&2:1&2:3&2:1&2:3&1:3 \\
\hline
4& & &1&1&1&1&1&1&1&2:2&2:2&2:4&2:2&2:4&2:4 \\
\hline
5& &1& & &1&1& &1& & & & & & & \\
\hline
6& &1& &1&1&1&1&1&1&3:3&3:4&3:2&3:4&3:2&4:2 \\
\hline
7& &1&1& &1&1& &1& & & & & & & \\
\hline
8& &1&1&1&1&1&1&1&1&4:4&4:4&4:4&4:4&4:4&4:4 \\
\hline
9&1& & & &1&1&1&1&1&5:5&5:5&5:5&5:5&5:5&5:5 \\
\hline
10&1& & &1&1&1&1&1&1&5:5&5:6&5:6&5:6&5:6&6:6 \\
\hline
11&1& &1& &1&1& &1& & & & & & & \\
\hline
12&1& &1&1& &1& & & & & & & & & \\
\hline
13&1&1& & &1&1&1&1&1&7:7&7:7&7:5&7:7&7:5&7:5 \\
\hline
14&1&1& &1&1&1&1&1&1&7:7&7:8&7:6&7:8&7:6&8:6 \\
\hline
15&1&1&1& &1&1&1&1&1&8:8&8:7&8:7&8:7&8:7&7:7 \\
\hline
16&1&1&1&1&1&1&1& & & & & & & & \\
\hline\hline
\end{array}
\]

\section{Example 2}
The 3-SAT instance from \cite{mail}:
\[
f = (p \vee q \vee r) \wedge (p \vee q \vee \bar{r}) \wedge (\bar{p} \vee s) \wedge (\bar{p} \vee \bar{s}) \wedge \bar{q}.
\]
Let's use the formula (4) from \cite{gubin} to iterate the compatibility matrices. 
\newline\indent
For example, compatibility matrices $C_{14,0} = c_{1} : c_{4}, ~ C_{15,0} = c_{1}:c_{5},$ and $C_{45,0} = c_{4}:c_{5}$ are (see below) 
\[
C_{14,0} = 
\begin{array}{|c|c|c|c|}
\hline
 & & &~  \\
\hline
1&1& &  \\
\hline
1&1& &  \\
\hline
1&1& &  \\
\hline
 & &1&  \\
\hline
 & &1&  \\
\hline
 & &1&  \\
\hline
 & &1&  \\
\hline
\end{array}
~~C_{15,0} = 
\begin{array}{|c|c|}
\hline
 &~  \\
\hline
1&  \\
\hline
 &  \\
\hline
 &  \\
\hline
1&  \\
\hline
1&  \\
\hline
 &  \\
\hline
 &  \\
\hline
\end{array}
~~C_{45,0} = 
\begin{array}{|c|c|}
\hline
1&~ \\
\hline
1&  \\
\hline
1&  \\
\hline
 &  \\
\hline
\end{array}
\]
Then, formula (4) from \cite{gubin} works in the following way:
\[
C_{45,1} = (C_{14,0}^{T} \times C_{15,0}) \wedge C_{45,0} =
\]
\[
= (~
\begin{array}{|c|c|c|c|c|c|c|c|}
\hline
~&1&1&1&~&~&~&~ \\
\hline
~&1&1&1&~&~&~&~ \\
\hline
 & & & &1&1&1&1 \\
\hline
~& & & &~&~&~&~ \\
\hline
\end{array}
\times
\begin{array}{|c|c|}
\hline
 & ~ \\
\hline
1&  \\
\hline
 &  \\
\hline
 &  \\
\hline
1&  \\
\hline
1&  \\
\hline
 &  \\
\hline
 &  \\
\hline
\end{array}
~)\wedge
\begin{array}{|c|c|}
\hline
1&~  \\
\hline
1&  \\
\hline
1&  \\
\hline
 &  \\
\hline
\end{array}
=
\]
\[
=
\begin{array}{|c|c|}
\hline
1&~  \\
\hline
1&  \\
\hline
1&  \\
\hline
 &  \\
\hline
\end{array}
\wedge
\begin{array}{|c|c|}
\hline
1&~  \\
\hline
1&  \\
\hline
1&  \\
\hline
 &  \\
\hline
\end{array}
=
\begin{array}{|c|c|}
\hline
1&~  \\
\hline
1&  \\
\hline
1&  \\
\hline
 &  \\
\hline
\end{array}.
\]
Multiplication of Boolean matrices can be defined in different ways. The rules used here are the rules of numeric matrices' multiplication but with replacement of numeric multiplications and sums with conjunctions and disjunctions appropriately \cite{gubin}.
\newline\indent
Let's return to the problem. Clauses' truth-tables:
\[
c_{1} = 
\begin{array}{||c||c|c|c||c||}
\hline\hline
\#&p&q&r&p \vee q \vee r \\
\hline\hline
1& & & &  \\
\hline
2& & &1&1 \\
\hline
3& &1& &1 \\
\hline
4& &1&1&1 \\
\hline
5&1& & &1 \\
\hline
6&1& &1&1 \\
\hline
7&1&1& &1 \\
\hline
8&1&1&1&1 \\
\hline\hline
\end{array}
~ ~
c_{2} = 
\begin{array}{||c||c|c|c||c||}
\hline\hline
\#&p&q&r&p \vee q \vee \bar{r} \\
\hline\hline
1& & & &1 \\
\hline
2& & &1&  \\
\hline
3& &1& &1 \\
\hline
4& &1&1&1 \\
\hline
5&1& & &1 \\
\hline
6&1& &1&1 \\
\hline
7&1&1& &1 \\
\hline
8&1&1&1&1 \\
\hline\hline
\end{array}
\]
\[
c_{3} = 
\begin{array}{||c||c|c||c||}
\hline\hline
\#&p&s&\bar{p} \vee s \\
\hline\hline
1& & &1 \\
\hline
2& &1&1 \\
\hline
3&1& &  \\
\hline
4&1&1&1 \\
\hline\hline
\end{array}
~ ~
c_{4} = 
\begin{array}{||c||c|c||c||}
\hline\hline
\#&p&s&\bar{p} \vee \bar{s} \\
\hline\hline
1& & &1 \\
\hline
2& &1&1 \\
\hline
3&1& &1 \\
\hline
4&1&1&  \\
\hline\hline
\end{array}
~~
c_{5} = 
\begin{array}{||c||c||c||}
\hline\hline
\#&q&\bar{q} \\
\hline\hline
1& &1 \\
\hline
2&1&  \\
\hline\hline
\end{array}
\]
\newpage
\emph{Start.} Compatibility matrices:
\[
\begin{array}{ccccc}
&c_{2}&c_{3}&c_{4}&c_{5} \\
c_{1}&
\begin{array}{||c||c|c|c|c|c|c|c|c||}
\hline\hline
&1&2&3&4&5&6&7&8 \\
\hline\hline
1& & & & & & & &  \\
\hline
2& & & & & & & &  \\
\hline
3& & &1& & & & &  \\
\hline
4& & & &1& & & &  \\
\hline
5& & & & &1& & &  \\
\hline
6& & & & & &1& &  \\
\hline
7& & & & & & &1&  \\
\hline
8& & & & & & & &1 \\
\hline\hline
\end{array}
&
\begin{array}{||c|c|c|c||}
\hline\hline
1&2&3&4 \\
\hline\hline
 & & &  \\
\hline
1&1& &  \\
\hline
1&1& &  \\
\hline
1&1& &  \\
\hline
 & & &1 \\
\hline
 & & &1 \\
\hline
 & & &1 \\
\hline
 & & &1 \\
\hline\hline
\end{array}
&
\begin{array}{||c|c|c|c||}
\hline\hline
1&2&3&4 \\
\hline\hline
 & & &  \\
\hline
1&1& &  \\
\hline
1&1& &  \\
\hline
1&1& &  \\
\hline
 & &1&  \\
\hline
 & &1&  \\
\hline
 & &1&  \\
\hline
 & &1&  \\
\hline\hline
\end{array}
&
\begin{array}{||c|c||}
\hline\hline
1&2 \\
\hline\hline
 &  \\
\hline
1&  \\
\hline
 &  \\
\hline
 &  \\
\hline
1&  \\
\hline
1&  \\
\hline
 &  \\
\hline
 &  \\
\hline\hline
\end{array}
\end{array}
\]

\[
\begin{array}{ccccc}
&&c_{3}&c_{4}&c_{5} \\
\begin{array}{cccccccc}
~~~~~~&~ &~ &~ &~ &~ &~ &~  \\
 & & & & & & &  \\
 & & & & & & &  \\
 & & & & & & &  \\
 & & & & & & &  \\
 & & & & & & &  \\
 & & & & & & &  \\
 & & & & & & &  \\
 & & & & & & &  \\
\end{array}
&
c_{2}&
\begin{array}{||c||c|c|c|c||}
\hline\hline
&1&2&3&4 \\
\hline\hline
1&1&1& &  \\
\hline
2& & & &  \\
\hline
3&1&1& &  \\
\hline
4&1&1& &  \\
\hline
5& & & &1 \\
\hline
6& & & &1 \\
\hline
7& & & &1 \\
\hline
8& & & &1 \\
\hline\hline
\end{array}
&
\begin{array}{||c|c|c|c||}
\hline\hline
1&2&3&4 \\
\hline\hline
1&1& &  \\
\hline
 & & &  \\
\hline
1&1& &  \\
\hline
1&1& &  \\
\hline
 & &1&  \\
\hline
 & &1&  \\
\hline
 & &1&  \\
\hline
 & &1&  \\
\hline\hline
\end{array}
&
\begin{array}{||c|c||}
\hline\hline
1&2 \\
\hline\hline
1&  \\
\hline
 &  \\
\hline
 &  \\
\hline
 &  \\
\hline
1&  \\
\hline
1&  \\
\hline
 &  \\
\hline
 &  \\
\hline\hline
\end{array}
\end{array}
\]

\[
\begin{array}{ccccc}
&&&c_{4}&c_{5} \\
\begin{array}{cccccccc}
~~~~~~&~ &~ &~ &~ &~ &~ &~  \\
 & & & & & & &  \\
 & & & & & & &  \\
 & & & & & & &  \\
 & & & & & & &  \\
\end{array}
&
\begin{array}{cccc}
~~~~~~&~ &~ &~  \\
 & & &  \\
 & & &  \\
 & & &  \\
 & & &  \\
\end{array}
&
c_{3}&
\begin{array}{||c||c|c|c|c||}
\hline\hline
&1&2&3&4 \\
\hline\hline
1&1& & &  \\
\hline
2& &1& &  \\
\hline
3& & & &  \\
\hline
4& & & &  \\
\hline\hline
\end{array}
&
\begin{array}{||c|c||}
\hline\hline
1&2 \\
\hline\hline
1&  \\
\hline
1&  \\
\hline
 &  \\
\hline
1&  \\
\hline\hline
\end{array}
\end{array}
\]

\[
\begin{array}{ccccc}
&&&&c_{5} \\
\begin{array}{cccccccc}
~~~~~~&~ &~ &~ &~ &~ &~ &~  \\
 & & & & & & &  \\
 & & & & & & &  \\
 & & & & & & &  \\
 & & & & & & &  \\
\end{array}
&
\begin{array}{cccc}
~~~~~&~ &~ &~  \\
 & & &  \\
 & & &  \\
 & & &  \\
 & & &  \\
\end{array}
&
\begin{array}{cccc}
~~~~~&~ &~ &~  \\
 & & &  \\
 & & &  \\
 & & &  \\
 & & &  \\
\end{array}
&
c_{4}&
\begin{array}{||c||c|c||}
\hline\hline
&1&2 \\
\hline\hline
1&1&  \\
\hline
2&1&  \\
\hline
3&1&  \\
\hline
4& &  \\
\hline\hline
\end{array}
\end{array}
\]

\newpage
\emph{Step 1.} Compatibility matrices depleted by clause $c_{1}$ with formula (4) from \cite{gubin}:
\[
\begin{array}{ccccc}
&c_{2}&c_{3}&c_{4}&c_{5} \\
c_{1}&
\begin{array}{||c||c|c|c|c|c|c|c|c||}
\hline\hline
&1&2&3&4&5&6&7&8 \\
\hline\hline
1& & & & & & & &  \\
\hline
2& & & & & & & &  \\
\hline
3& & &1& & & & &  \\
\hline
4& & & &1& & & &  \\
\hline
5& & & & &1& & &  \\
\hline
6& & & & & &1& &  \\
\hline
7& & & & & & &1&  \\
\hline
8& & & & & & & &1 \\
\hline\hline
\end{array}
&
\begin{array}{||c|c|c|c||}
\hline\hline
1&2&3&4 \\
\hline\hline
 & & &  \\
\hline
1&1& &  \\
\hline
1&1& &  \\
\hline
1&1& &  \\
\hline
 & & &1 \\
\hline
 & & &1 \\
\hline
 & & &1 \\
\hline
 & & &1 \\
\hline\hline
\end{array}
&
\begin{array}{||c|c|c|c||}
\hline\hline
1&2&3&4 \\
\hline\hline
 & & &  \\
\hline
1&1& &  \\
\hline
1&1& &  \\
\hline
1&1& &  \\
\hline
 & &1&  \\
\hline
 & &1&  \\
\hline
 & &1&  \\
\hline
 & &1&  \\
\hline\hline
\end{array}
&
\begin{array}{||c|c||}
\hline\hline
1&2 \\
\hline\hline
 &  \\
\hline
1&  \\
\hline
 &  \\
\hline
 &  \\
\hline
1&  \\
\hline
1&  \\
\hline
 &  \\
\hline
 &  \\
\hline\hline
\end{array}
\end{array}
\]

\[
\begin{array}{ccccc}
&&c_{3}&c_{4}&c_{5} \\
\begin{array}{cccccccc}
~~~~~~&~ &~ &~ &~ &~ &~ &~  \\
 & & & & & & &  \\
 & & & & & & &  \\
 & & & & & & &  \\
 & & & & & & &  \\
 & & & & & & &  \\
 & & & & & & &  \\
 & & & & & & &  \\
 & & & & & & &  \\
\end{array}
&
c_{2}&
\begin{array}{||c||c|c|c|c||}
\hline\hline
&1&2&3&4 \\
\hline\hline
1& & & &  \\
\hline
2& & & &  \\
\hline
3&1&1& &  \\
\hline
4&1&1& &  \\
\hline
5& & & &1 \\
\hline
6& & & &1 \\
\hline
7& & & &1 \\
\hline
8& & & &1 \\
\hline\hline
\end{array}
&
\begin{array}{||c|c|c|c||}
\hline\hline
1&2&3&4 \\
\hline\hline
 & & &  \\
\hline
 & & &  \\
\hline
1&1& &  \\
\hline
1&1& &  \\
\hline
 & &1&  \\
\hline
 & &1&  \\
\hline
 & &1&  \\
\hline
 & &1&  \\
\hline\hline
\end{array}
&
\begin{array}{||c|c||}
\hline\hline
1&2 \\
\hline\hline
 &  \\
\hline
 &  \\
\hline
 &  \\
\hline
 &  \\
\hline
1&  \\
\hline
1&  \\
\hline
 &  \\
\hline
 &  \\
\hline\hline
\end{array}
\end{array}
\]

\[
\begin{array}{ccccc}
&&&c_{4}&c_{5} \\
\begin{array}{cccccccc}
~~~~~~&~ &~ &~ &~ &~ &~ &~  \\
 & & & & & & &  \\
 & & & & & & &  \\
 & & & & & & &  \\
 & & & & & & &  \\
\end{array}
&
\begin{array}{cccc}
~~~~~~&~ &~ &~  \\
 & & &  \\
 & & &  \\
 & & &  \\
 & & &  \\
\end{array}
&
c_{3}&
\begin{array}{||c||c|c|c|c||}
\hline\hline
&1&2&3&4 \\
\hline\hline
1&1& & &  \\
\hline
2& &1& &  \\
\hline
3& & & &  \\
\hline
4& & & &  \\
\hline\hline
\end{array}
&
\begin{array}{||c|c||}
\hline\hline
1&2 \\
\hline\hline
1&  \\
\hline
1&  \\
\hline
 &  \\
\hline
1&  \\
\hline\hline
\end{array}
\end{array}
\]

\[
\begin{array}{ccccc}
&&&&c_{5} \\
\begin{array}{cccccccc}
~~~~~~&~ &~ &~ &~ &~ &~ &~  \\
 & & & & & & &  \\
 & & & & & & &  \\
 & & & & & & &  \\
 & & & & & & &  \\
\end{array}
&
\begin{array}{cccc}
~~~~~&~ &~ &~  \\
 & & &  \\
 & & &  \\
 & & &  \\
 & & &  \\
\end{array}
&
\begin{array}{cccc}
~~~~~&~ &~ &~  \\
 & & &  \\
 & & &  \\
 & & &  \\
 & & &  \\
\end{array}
&
c_{4}&
\begin{array}{||c||c|c||}
\hline\hline
&1&2 \\
\hline\hline
1&1&  \\
\hline
2&1&  \\
\hline
3&1&  \\
\hline
4& &  \\
\hline\hline
\end{array}
\end{array}
\]

\newpage
\emph{Step 2.} Compatibility matrices depleted by clause $c_{2}$ with formula (4) from \cite{gubin}:
\[
\begin{array}{ccccc}
&c_{2}&c_{3}&c_{4}&c_{5} \\
c_{1}&
\begin{array}{||c||c|c|c|c|c|c|c|c||}
\hline\hline
&1&2&3&4&5&6&7&8 \\
\hline\hline
1& & & & & & & &  \\
\hline
2& & & & & & & &  \\
\hline
3& & &1& & & & &  \\
\hline
4& & & &1& & & &  \\
\hline
5& & & & &1& & &  \\
\hline
6& & & & & &1& &  \\
\hline
7& & & & & & &1&  \\
\hline
8& & & & & & & &1 \\
\hline\hline
\end{array}
&
\begin{array}{||c|c|c|c||}
\hline\hline
1&2&3&4 \\
\hline\hline
 & & &  \\
\hline
1&1& &  \\
\hline
1&1& &  \\
\hline
1&1& &  \\
\hline
 & & &1 \\
\hline
 & & &1 \\
\hline
 & & &1 \\
\hline
 & & &1 \\
\hline\hline
\end{array}
&
\begin{array}{||c|c|c|c||}
\hline\hline
1&2&3&4 \\
\hline\hline
 & & &  \\
\hline
1&1& &  \\
\hline
1&1& &  \\
\hline
1&1& &  \\
\hline
 & &1&  \\
\hline
 & &1&  \\
\hline
 & &1&  \\
\hline
 & &1&  \\
\hline\hline
\end{array}
&
\begin{array}{||c|c||}
\hline\hline
1&2 \\
\hline\hline
 &  \\
\hline
1&  \\
\hline
 &  \\
\hline
 &  \\
\hline
1&  \\
\hline
1&  \\
\hline
 &  \\
\hline
 &  \\
\hline\hline
\end{array}
\end{array}
\]

\[
\begin{array}{ccccc}
&&c_{3}&c_{4}&c_{5} \\
\begin{array}{cccccccc}
~~~~~~&~ &~ &~ &~ &~ &~ &~  \\
 & & & & & & &  \\
 & & & & & & &  \\
 & & & & & & &  \\
 & & & & & & &  \\
 & & & & & & &  \\
 & & & & & & &  \\
 & & & & & & &  \\
 & & & & & & &  \\
\end{array}
&
c_{2}&
\begin{array}{||c||c|c|c|c||}
\hline\hline
&1&2&3&4 \\
\hline\hline
1& & & &  \\
\hline
2& & & &  \\
\hline
3&1&1& &  \\
\hline
4&1&1& &  \\
\hline
5& & & &1 \\
\hline
6& & & &1 \\
\hline
7& & & &1 \\
\hline
8& & & &1 \\
\hline\hline
\end{array}
&
\begin{array}{||c|c|c|c||}
\hline\hline
1&2&3&4 \\
\hline\hline
 & & &  \\
\hline
 & & &  \\
\hline
1&1& &  \\
\hline
1&1& &  \\
\hline
 & &1&  \\
\hline
 & &1&  \\
\hline
 & &1&  \\
\hline
 & &1&  \\
\hline\hline
\end{array}
&
\begin{array}{||c|c||}
\hline\hline
1&2 \\
\hline\hline
 &  \\
\hline
 &  \\
\hline
 &  \\
\hline
 &  \\
\hline
1&  \\
\hline
1&  \\
\hline
 &  \\
\hline
 &  \\
\hline\hline
\end{array}
\end{array}
\]

\[
\begin{array}{ccccc}
&&&c_{4}&c_{5} \\
\begin{array}{cccccccc}
~~~~~~&~ &~ &~ &~ &~ &~ &~  \\
 & & & & & & &  \\
 & & & & & & &  \\
 & & & & & & &  \\
 & & & & & & &  \\
\end{array}
&
\begin{array}{cccc}
~~~~~~&~ &~ &~  \\
 & & &  \\
 & & &  \\
 & & &  \\
 & & &  \\
\end{array}
&
c_{3}&
\begin{array}{||c||c|c|c|c||}
\hline\hline
&1&2&3&4 \\
\hline\hline
1&1& & &  \\
\hline
2& &1& &  \\
\hline
3& & & &  \\
\hline
4& & & &  \\
\hline\hline
\end{array}
&
\begin{array}{||c|c||}
\hline\hline
1&2 \\
\hline\hline
 &  \\
\hline
 &  \\
\hline
 &  \\
\hline
1&  \\
\hline\hline
\end{array}
\end{array}
\]

\[
\begin{array}{ccccc}
&&&&c_{5} \\
\begin{array}{cccccccc}
~~~~~~&~ &~ &~ &~ &~ &~ &~  \\
 & & & & & & &  \\
 & & & & & & &  \\
 & & & & & & &  \\
 & & & & & & &  \\
\end{array}
&
\begin{array}{cccc}
~~~~~&~ &~ &~  \\
 & & &  \\
 & & &  \\
 & & &  \\
 & & &  \\
\end{array}
&
\begin{array}{cccc}
~~~~~&~ &~ &~  \\
 & & &  \\
 & & &  \\
 & & &  \\
 & & &  \\
\end{array}
&
c_{4}&
\begin{array}{||c||c|c||}
\hline\hline
&1&2 \\
\hline\hline
1& &  \\
\hline
2& &  \\
\hline
3&1&  \\
\hline
4& &  \\
\hline\hline
\end{array}
\end{array}
\]

\newpage
\emph{Step 3.} Compatibility matrices depleted by clause $c_{3}$ with formula (4) from \cite{gubin}:
\[
\begin{array}{ccccc}
&c_{2}&c_{3}&c_{4}&c_{5} \\
c_{1}&
\begin{array}{||c||c|c|c|c|c|c|c|c||}
\hline\hline
&1&2&3&4&5&6&7&8 \\
\hline\hline
1& & & & & & & &  \\
\hline
2& & & & & & & &  \\
\hline
3& & &1& & & & &  \\
\hline
4& & & &1& & & &  \\
\hline
5& & & & &1& & &  \\
\hline
6& & & & & &1& &  \\
\hline
7& & & & & & &1&  \\
\hline
8& & & & & & & &1 \\
\hline\hline
\end{array}
&
\begin{array}{||c|c|c|c||}
\hline\hline
1&2&3&4 \\
\hline\hline
 & & &  \\
\hline
1&1& &  \\
\hline
1&1& &  \\
\hline
1&1& &  \\
\hline
 & & &1 \\
\hline
 & & &1 \\
\hline
 & & &1 \\
\hline
 & & &1 \\
\hline\hline
\end{array}
&
\begin{array}{||c|c|c|c||}
\hline\hline
1&2&3&4 \\
\hline\hline
 & & &  \\
\hline
1&1& &  \\
\hline
1&1& &  \\
\hline
1&1& &  \\
\hline
 & &1&  \\
\hline
 & &1&  \\
\hline
 & &1&  \\
\hline
 & &1&  \\
\hline\hline
\end{array}
&
\begin{array}{||c|c||}
\hline\hline
1&2 \\
\hline\hline
 &  \\
\hline
1&  \\
\hline
 &  \\
\hline
 &  \\
\hline
1&  \\
\hline
1&  \\
\hline
 &  \\
\hline
 &  \\
\hline\hline
\end{array}
\end{array}
\]

\[
\begin{array}{ccccc}
&&c_{3}&c_{4}&c_{5} \\
\begin{array}{cccccccc}
~~~~~~&~ &~ &~ &~ &~ &~ &~  \\
 & & & & & & &  \\
 & & & & & & &  \\
 & & & & & & &  \\
 & & & & & & &  \\
 & & & & & & &  \\
 & & & & & & &  \\
 & & & & & & &  \\
 & & & & & & &  \\
\end{array}
&
c_{2}&
\begin{array}{||c||c|c|c|c||}
\hline\hline
&1&2&3&4 \\
\hline\hline
1& & & &  \\
\hline
2& & & &  \\
\hline
3&1&1& &  \\
\hline
4&1&1& &  \\
\hline
5& & & &1 \\
\hline
6& & & &1 \\
\hline
7& & & &1 \\
\hline
8& & & &1 \\
\hline\hline
\end{array}
&
\begin{array}{||c|c|c|c||}
\hline\hline
1&2&3&4 \\
\hline\hline
 & & &  \\
\hline
 & & &  \\
\hline
1&1& &  \\
\hline
1&1& &  \\
\hline
 & &1&  \\
\hline
 & &1&  \\
\hline
 & &1&  \\
\hline
 & &1&  \\
\hline\hline
\end{array}
&
\begin{array}{||c|c||}
\hline\hline
1&2 \\
\hline\hline
 &  \\
\hline
 &  \\
\hline
 &  \\
\hline
 &  \\
\hline
1&  \\
\hline
1&  \\
\hline
 &  \\
\hline
 &  \\
\hline\hline
\end{array}
\end{array}
\]

\[
\begin{array}{ccccc}
&&&c_{4}&c_{5} \\
\begin{array}{cccccccc}
~~~~~~&~ &~ &~ &~ &~ &~ &~  \\
 & & & & & & &  \\
 & & & & & & &  \\
 & & & & & & &  \\
 & & & & & & &  \\
\end{array}
&
\begin{array}{cccc}
~~~~~~&~ &~ &~  \\
 & & &  \\
 & & &  \\
 & & &  \\
 & & &  \\
\end{array}
&
c_{3}&
\begin{array}{||c||c|c|c|c||}
\hline\hline
&1&2&3&4 \\
\hline\hline
1&1& & &  \\
\hline
2& &1& &  \\
\hline
3& & & &  \\
\hline
4& & & &  \\
\hline\hline
\end{array}
&
\begin{array}{||c|c||}
\hline\hline
1&2 \\
\hline\hline
 &  \\
\hline
 &  \\
\hline
 &  \\
\hline
1&  \\
\hline\hline
\end{array}
\end{array}
\]

\[
\begin{array}{ccccc}
&&&&c_{5} \\
\begin{array}{cccccccc}
~~~~~~&~ &~ &~ &~ &~ &~ &~  \\
 & & & & & & &  \\
 & & & & & & &  \\
 & & & & & & &  \\
 & & & & & & &  \\
\end{array}
&
\begin{array}{cccc}
~~~~~&~ &~ &~  \\
 & & &  \\
 & & &  \\
 & & &  \\
 & & &  \\
\end{array}
&
\begin{array}{cccc}
~~~~~&~ &~ &~  \\
 & & &  \\
 & & &  \\
 & & &  \\
 & & &  \\
\end{array}
&
c_{4}&
\begin{array}{||c||c|c||}
\hline\hline
&1&2 \\
\hline\hline
1& &  \\
\hline
2& &  \\
\hline
3& &  \\
\hline
4& &  \\
\hline\hline
\end{array}
\end{array}
\]
The compatibility matrix $C_{45,3} = c_{4} : c_{5}$ is a $false$-matrix. Thus, the given 3-SAT instance is unsatisfiable.
\newline\indent
Let's emphasize that initial compatibility matrix $c_{4} : c_{5}$ was depleted by clauses $c_{4}$ and $c_{5}$. It contained (its $true$-elements were associated with) only compatible true assignments of clauses $c_{4}$ and $c_{5}$. The algorithm's steps 1, 2, and 3 cleaned those true assignments to make them compatible with clauses $c_{1}, ~ c_{2},$ and $c_{3}$, also. It so happened that there were no possibilities left after that cleaning in this example.
\newline\indent
The tree of possibilities collapses due to formula (4) \cite{gubin}. The Boolean matrices' multiplication ``totals'' the possibilities. And the Boolean matrices' conjunction ``clears'' the totals. Sure, some totals can be empty, too.
\newline\indent
For example, let's take $(3,1)$-element of matrix $C_{45,3}$ (Step 3). The multiplication part of formula (4) \cite{gubin} totals all possibilities sorted over the third columns of matrix $C_{34,2}$ and the first column of matrix $C_{35,2}$ into only one possibility: 
\[
false \wedge false \vee false \wedge false \vee false \wedge false \vee false \wedge true = false.
\]
That means that there are no possibilities left. Then, the conjunction part of the formula (4) \cite{gubin} clears the total possibility against possibilities already accounted for in the $(3,1)$-element of matrix $C_{45,2}$: 
\[
false \wedge true = false.
\]

\section{Example 3}
Let's see how the algorithm deals with 2-SAT. The 2-SAT instance from \cite{contra}:
\[
f = \bar{p} \wedge \bar{q} \wedge \bar{r} \wedge (p \vee q) \wedge (p \vee r) \wedge (q \vee r).
\]
Let's use formula (4) from \cite{gubin} to iterate the compatibility matrices. 
\newline\indent
Clauses' truth-tables:
\[
c_{1} = 
\begin{array}{||c||c||c||}
\hline\hline
\#&p&\bar{p} \\
\hline\hline
1& &1 \\
\hline
2&1&  \\
\hline\hline
\end{array}
~ ~
c_{2} = 
\begin{array}{||c||c||c||}
\hline\hline
\#&q&\bar{q} \\
\hline\hline
1& &1 \\
\hline
2&1&  \\
\hline\hline
\end{array}
~~
c_{3} = 
\begin{array}{||c||c||c||}
\hline\hline
\#&r&\bar{r} \\
\hline\hline
1& &1 \\
\hline
2&1&  \\
\hline\hline
\end{array}
\]
\[
c_{4} = 
\begin{array}{||c||c|c||c||}
\hline\hline
\#&p&q&p \vee q \\
\hline\hline
1& & &  \\
\hline
2& &1&1 \\
\hline
3&1& &1 \\
\hline
4&1&1&1 \\
\hline\hline
\end{array}
~ ~
c_{5} = 
\begin{array}{||c||c|c||c||}
\hline\hline
\#&p&r&p \vee r \\
\hline\hline
1& & &  \\
\hline
2& &1&1 \\
\hline
3&1& &1 \\
\hline
4&1&1&1 \\
\hline\hline
\end{array}
~~
c_{6} = 
\begin{array}{||c||c|c||c||}
\hline\hline
\#&q&r&q \vee r \\
\hline\hline
1& & &  \\
\hline
2& &1&1 \\
\hline
3&1& &1 \\
\hline
4&1&1&1 \\
\hline\hline
\end{array}
\]
\newpage
\emph{Start.} Compatibility matrices:
\[
\begin{array}{cccccc}
&c_{2}&c_{3}&c_{4}&c_{5}&c_{6} \\
c_{1}&
\begin{array}{||c||c|c||}
\hline\hline
&1&2 \\
\hline\hline
1&1&  \\
\hline
2& &  \\
\hline\hline
\end{array}
&
\begin{array}{||c|c||}
\hline\hline
1&2 \\
\hline\hline
1&  \\
\hline
 &  \\
\hline\hline
\end{array}
&
\begin{array}{||c|c|c|c||}
\hline\hline
1&2&3&4 \\
\hline\hline
 &1& &  \\
\hline
 & & &  \\
\hline\hline
\end{array}
&
\begin{array}{||c|c|c|c||}
\hline\hline
1&2&3&4 \\
\hline\hline
 &1& &  \\
\hline
 & & &  \\
\hline\hline
\end{array}
&
\begin{array}{||c|c|c|c||}
\hline\hline
1&2&3&4 \\
\hline\hline
 &1&1&1 \\
\hline
 & & &  \\
\hline\hline
\end{array}
\end{array}
\]

\[
\begin{array}{cccccc}
&&c_{3}&c_{4}&c_{5}&c_{6} \\
\begin{array}{cc}
~&~~~~ \\
~&~ \\
~&~ \\
\end{array}
&c_{2}&
\begin{array}{||c||c|c||}
\hline\hline
&1&2 \\
\hline\hline
1&1&  \\
\hline
2& &  \\
\hline\hline
\end{array}
&
\begin{array}{||c|c|c|c||}
\hline\hline
1&2&3&4 \\
\hline\hline
 & &1&  \\
\hline
 & & &  \\
\hline\hline
\end{array}
&
\begin{array}{||c|c|c|c||}
\hline\hline
1&2&3&4 \\
\hline\hline
 &1&1&1 \\
\hline
 & & &  \\
\hline\hline
\end{array}
&
\begin{array}{||c|c|c|c||}
\hline\hline
1&2&3&4 \\
\hline\hline
 &1&&  \\
\hline
 & & &  \\
\hline\hline
\end{array}
\end{array}
\]

\[
\begin{array}{cccccc}
&&&c_{4}&c_{5}&c_{6} \\
\begin{array}{cc}
~&~~~~ \\
~&~ \\
~&~ \\
\end{array}
&
\begin{array}{cc}
~&~~~~ \\
~&~ \\
~&~ \\
\end{array}
&c_{3}&
\begin{array}{||c||c|c|c|c||}
\hline\hline
&1&2&3&4 \\
\hline\hline
1& &1&1&1 \\
\hline
2& & & &  \\
\hline\hline
\end{array}
&
\begin{array}{||c|c|c|c||}
\hline\hline
1&2&3&4 \\
\hline\hline
 & &1&  \\
\hline
 & & &  \\
\hline\hline
\end{array}
&
\begin{array}{||c|c|c|c||}
\hline\hline
1&2&3&4 \\
\hline\hline
 & &1&  \\
\hline
 & & &  \\
\hline\hline
\end{array}
\end{array}
\]

\[
\begin{array}{cccccc}
&&&&c_{5}&c_{6} \\
\begin{array}{cc}
~&~~~~ \\
~&~ \\
~&~ \\
\end{array}
&
\begin{array}{cc}
~&~~~~ \\
~&~ \\
~&~ \\
\end{array}
&
\begin{array}{cccc}
~&~&~&~~~~~ \\
 & & &  \\
 & & &  \\
\end{array}
&c_{4}&
\begin{array}{||c||c|c|c|c||}
\hline\hline
&1&2&3&4 \\
\hline\hline
1& & & &  \\
\hline
2& &1& &  \\
\hline
3& & &1&1 \\
\hline
4& & &1&1 \\
\hline\hline
\end{array}
&
\begin{array}{||c|c|c|c||}
\hline\hline
1&2&3&4 \\
\hline\hline
 & & &  \\
\hline
 & &1&1 \\
\hline
 &1& &  \\
\hline
 & &1&1 \\
\hline\hline
\end{array}
\end{array}
\]

\[
\begin{array}{cccccc}
&&&&&c_{6} \\
\begin{array}{cc}
~&~~~~ \\
~&~ \\
~&~ \\
\end{array}
&
\begin{array}{cc}
~&~~~~ \\
~&~ \\
~&~ \\
\end{array}
&
\begin{array}{cccc}
~&~&~&~~~~~ \\
 & & &  \\
 & & &  \\
\end{array}
&
\begin{array}{cccc}
~&~&~&~~~~~ \\
 & & &  \\
 & & &  \\
\end{array}
&c_{5}&
\begin{array}{||c||c|c|c|c||}
\hline\hline
&1&2&3&4 \\
\hline\hline
1& & & &  \\
\hline
2& &1& &1 \\
\hline
3& & &1&  \\
\hline
4& &1& &1 \\
\hline\hline
\end{array}
\end{array}
\]

\newpage
\emph{Step 1.} Compatibility matrices depleted by clause $c_{1}$ with formula (4) from \cite{gubin}:
\[
\begin{array}{cccccc}
&c_{2}&c_{3}&c_{4}&c_{5}&c_{6} \\
c_{1}&
\begin{array}{||c||c|c||}
\hline\hline
&1&2 \\
\hline\hline
1&1&  \\
\hline
2& &  \\
\hline\hline
\end{array}
&
\begin{array}{||c|c||}
\hline\hline
1&2 \\
\hline\hline
1&  \\
\hline
 &  \\
\hline\hline
\end{array}
&
\begin{array}{||c|c|c|c||}
\hline\hline
1&2&3&4 \\
\hline\hline
 &1& &  \\
\hline
 & & &  \\
\hline\hline
\end{array}
&
\begin{array}{||c|c|c|c||}
\hline\hline
1&2&3&4 \\
\hline\hline
 &1& &  \\
\hline
 & & &  \\
\hline\hline
\end{array}
&
\begin{array}{||c|c|c|c||}
\hline\hline
1&2&3&4 \\
\hline\hline
 &1&1&1 \\
\hline
 & & &  \\
\hline\hline
\end{array}
\end{array}
\]

\[
\begin{array}{cccccc}
&&c_{3}&c_{4}&c_{5}&c_{6} \\
\begin{array}{cc}
~&~~~~ \\
~&~ \\
~&~ \\
\end{array}
&c_{2}&
\begin{array}{||c||c|c||}
\hline\hline
&1&2 \\
\hline\hline
1&1&  \\
\hline
2& &  \\
\hline\hline
\end{array}
&
\begin{array}{||c|c|c|c||}
\hline\hline
1&2&3&4 \\
\hline\hline
 & & &  \\
\hline
 & & &  \\
\hline\hline
\end{array}
&
\begin{array}{||c|c|c|c||}
\hline\hline
1&2&3&4 \\
\hline\hline
 &1& &  \\
\hline
 & & &  \\
\hline\hline
\end{array}
&
\begin{array}{||c|c|c|c||}
\hline\hline
1&2&3&4 \\
\hline\hline
 &1& &  \\
\hline
 & & &  \\
\hline\hline
\end{array}
\end{array}
\]

\[
\begin{array}{cccccc}
&&&c_{4}&c_{5}&c_{6} \\
\begin{array}{cc}
~&~~~~ \\
~&~ \\
~&~ \\
\end{array}
&
\begin{array}{cc}
~&~~~~ \\
~&~ \\
~&~ \\
\end{array}
&c_{3}&
\begin{array}{||c||c|c|c|c||}
\hline\hline
&1&2&3&4 \\
\hline\hline
1& &1& &  \\
\hline
2& & & &  \\
\hline\hline
\end{array}
&
\begin{array}{||c|c|c|c||}
\hline\hline
1&2&3&4 \\
\hline\hline
 & & &  \\
\hline
 & & &  \\
\hline\hline
\end{array}
&
\begin{array}{||c|c|c|c||}
\hline\hline
1&2&3&4 \\
\hline\hline
 & &1&  \\
\hline
 & & &  \\
\hline\hline
\end{array}
\end{array}
\]

\[
\begin{array}{cccccc}
&&&&c_{5}&c_{6} \\
\begin{array}{cc}
~&~~~~ \\
~&~ \\
~&~ \\
\end{array}
&
\begin{array}{cc}
~&~~~~ \\
~&~ \\
~&~ \\
\end{array}
&
\begin{array}{cccc}
~&~&~&~~~~~ \\
 & & &  \\
 & & &  \\
\end{array}
&c_{4}&
\begin{array}{||c||c|c|c|c||}
\hline\hline
&1&2&3&4 \\
\hline\hline
1& & & &  \\
\hline
2& &1& &  \\
\hline
3& & & &  \\
\hline
4& & & &  \\
\hline\hline
\end{array}
&
\begin{array}{||c|c|c|c||}
\hline\hline
1&2&3&4 \\
\hline\hline
 & & &  \\
\hline
 & &1&1 \\
\hline
 & & &  \\
\hline
 & & &  \\
\hline\hline
\end{array}
\end{array}
\]

\[
\begin{array}{cccccc}
&&&&&c_{6} \\
\begin{array}{cc}
~&~~~~ \\
~&~ \\
~&~ \\
\end{array}
&
\begin{array}{cc}
~&~~~~ \\
~&~ \\
~&~ \\
\end{array}
&
\begin{array}{cccc}
~&~&~&~~~~~ \\
 & & &  \\
 & & &  \\
\end{array}
&
\begin{array}{cccc}
~&~&~&~~~~~ \\
 & & &  \\
 & & &  \\
\end{array}
&c_{5}&
\begin{array}{||c||c|c|c|c||}
\hline\hline
&1&2&3&4 \\
\hline\hline
1& & & &  \\
\hline
2& &1& &1 \\
\hline
3& & & &  \\
\hline
4& & & &  \\
\hline\hline
\end{array}
\end{array}
\]
Here, patterns of unsatisfiability arose - matrices $c_{2} : c_{4}$ and $c_{3} : c_{5}$ are $false$-matrices. Thus, the 2-SAT instance is unsatisfiable.
\newline\indent
Let's mention that $false$-matrix $c_{2} : c_{4}$ in Step 1 (matrix $C_{24,1}$) shows that formula $f$ contains an unsatisfiable part
\[
c_{1} \wedge c_{2} \wedge c_{4} = \bar{p} \wedge \bar{q} \wedge (p \vee q).
\]
And $false$-matrix $c_{3} : c_{5}$ in the Step 1 (matrix $C_{35,1}$) shows that formula $f$ contains an unsatisfiable part
\[
c_{1} \wedge c_{3} \wedge c_{5} = \bar{p} \wedge \bar{r} \wedge (p \vee r).
\]
Formula $f$ contains one more unsatisfiable part
\[
c_{2} \wedge c_{3} \wedge c_{6} = \bar{q} \wedge \bar{r} \wedge (q \vee r).
\]
But that would be detected only in Step 2 after the depletion of compatibility matrix $c_{3} : c_{6}$ by clause $c_{2}$.
\newline\indent
The phenomena noticed here are not accidental.
\begin{thm}(Structure of unsatisfiable SAT instances)
\newline
1). If the algorithm stops after $k$ steps, then the given propositional formula contains $k+2$ clauses that create an unsatisfiable formula.
\newline
2). If the given propositional formula contains $k$ clauses that create an unsatisfiable formula, then there is such an enumeration of clauses that the algorithm will stop in $k-2$ steps.
\end{thm}
For example, the algorithm will not even iterate for the formula
\[
x \wedge \bar{x}.
\]
It will detect that the formula is unsatisfiable on its initial step ``Start''.

\section{Example 4}
The following 3-SAT instances:
\[
f_{1} = (p \vee q \vee r) \wedge (p \vee q \vee \bar{r}) \wedge (p \vee \bar{q} \vee r) \wedge (p \vee \bar{q} \vee \bar{r}) \wedge (\bar{p} \vee q \vee r) \wedge (\bar{p} \vee q \vee \bar{r}) \wedge (\bar{p} \vee \bar{q} \vee r),
\]
\[
f_{2} = f_{1} \wedge (\bar{p} \vee \bar{q} \vee \bar{r}).
\]
Clauses and their truth-tables:
\[
\begin{array}{||c|c||}
\hline\hline
&c_{i} \\
\hline\hline
c_{1}&p \vee q \vee r \\
\hline
c_{2}&p \vee q \vee \bar{r} \\
\hline
c_{3}&p \vee \bar{q} \vee r \\
\hline
c_{4}&p \vee \bar{q} \vee \bar{r} \\
\hline
c_{5}&\bar{p} \vee q \vee r \\
\hline
c_{6}&\bar{p} \vee q \vee \bar{r} \\
\hline
c_{7}&\bar{p} \vee \bar{q} \vee r \\
\hline
c_{8}&\bar{p} \vee \bar{q} \vee \bar{r} \\
\hline\hline
\end{array}
~~
\begin{array}{||c||c|c|c||c|c|c|c|c|c|c|c||}
\hline\hline
\#&p&q&r&c_{1}&c_{2}&c_{3}&c_{4}&c_{5}&c_{6}&c_{7}&c_{8} \\
\hline\hline
1& & & & &1&1&1&1&1&1&1 \\
\hline
2& & &1&1& &1&1&1&1&1&1 \\
\hline
3& &1& &1&1& &1&1&1&1&1 \\
\hline
4& &1&1&1&1&1& &1&1&1&1 \\
\hline
5&1& & &1&1&1&1& &1&1&1 \\
\hline
6&1& &1&1&1&1&1&1& &1&1 \\
\hline
7&1&1& &1&1&1&1&1&1& &1 \\
\hline
8&1&1&1&1&1&1&1&1&1&1&  \\
\hline\hline
\end{array}
\]
Due to the selected enumeration of the strings in the truth-tables, the compatibility matrices are the following diagonal Boolean matrices:
\[
C_{ij} = (c_{i} \wedge c_{j}) = diag(b_{ij1},b_{ij2},b_{ij3},b_{ij4},b_{ij5},b_{ij6},b_{7},b_{ij8}),
\]
- where $i=1,2,\ldots,7;~j=2,3,\ldots,8;$ and 
\[
b_{ijk} = \left \{ \begin{array}{ll}
false&k = i \vee k = j \\
true&k \neq i \wedge k \neq j
\end{array}, \right .
\]
- where $k = 1,2,\ldots,8$. Thus, each iteration of the algorithm will eliminate (make it $false$) from the initial compatibility matrices at least one diagonal element. After $n \leq 8-2=6$ iterations, the result will contain only those elements, which pertain their value $true$. Those elements are common for all compatibility matrices. For formula $f_{1}$, such element is 
\[
p = q = r = true.
\]
But for formula $f_{2}$, there are no such elements. Thus, $f_{1}$ is satisfiable, and $f_{2}$ is not.

\end{document}